\documentclass[12pt]{article}
\begin{document}
\baselineskip=20pt

\title{Field spectrum and degrees of freedom in  AdS/CFT correspondence
and Randall Sundrum model}

\author{\large Henrique Boschi-Filho\footnote{\noindent e-mail: 
boschi @ if.ufrj.br}\,  
and 
Nelson R. F. Braga\footnote{\noindent e-mail: braga @ if.ufrj.br}
\\ 
\\ 
\it Instituto de F\'\i sica, Universidade Federal do Rio de Janeiro\\
\it Caixa Postal 68528, 21945-970  Rio de Janeiro, RJ, Brazil}
 
\date{}

\maketitle

\vskip 3cm

\begin{abstract}  
Compactified AdS space can not be mapped into just one Poincare 
coordinate chart. This implies that the  bulk field spectrum 
is discrete despite the infinite range of the  coordinates.
We discuss here why this discretization of the field spectrum 
seems to be a necessary  ingredient for the holographic mapping.
For the Randall Sundrum model we show that this discretization 
appears even without the second brane. 

\end{abstract}

%\vskip 3cm
%\noindent PACS: 03.70.+k, 04.20.Gz, 11.10.Kk .

\vfill\eject

\section{Introduction}

The interest of theoretical physicists in studying fields in
anti de Sitter (AdS) space is not new\cite{Fro}.
In particular the question of the quantization of fields in this space
circumventing the problem of the lack of a Cauchy surface
was addressed in\cite{QAdS1,QAdS2}. 
There was, however, a remarkable increase in the attention devoted
to this subject since the appearance of  two recent models where this 
geometry plays a special role.
The first, the so called AdS/CFT correspondence,  was motivated 
by the Maldacena \cite{Malda} conjecture on the equivalence 
(or duality) of the large $N$ limit of $SU(N)$ superconformal 
field theories in $n$ dimensions and supergravity and 
string theory in anti de Sitter spacetime in $n+1$  dimensions. 
This correspondence was elaborated by  
Gubser, Klebanov and Polyakov \cite{GKP} and  Witten \cite{Wi}
interpreting the boundary values of bulk fields as sources of 
boundary theory correlators.
The Maldacena conjecture and the subsequent 
work on AdS/CFT correspondence\cite{Malda2}  strongly reinforced the
relevance of understanding  the subtleties of field theories in 
AdS spaces. 

The second one, the Randall and Sundrum model\cite{RS1},
proposes  a solution to the hierarchy problem 
 between the mass scale of the standard model and the Plank scale. 
Their model is essentially defined in a slice of the five dimensional 
AdS space bounded by  two 3-branes and can  be regarded as an
alternative to usual compactification, as they discussed in  \cite{RS2}.
The standard model fields are confined to one of the branes while the
non factorizable form of the metric makes it possible to have propagation 
of gravity in the extra dimension without spoiling Newton's law 
up to experimental precision.
One important property of this model is that none of the two
branes is located at the AdS boundary. This way, it is possible to
include gravitational fluctuations on the branes\cite{Ver},
in contrast to the standard AdS/CFT scenario, where the singular form 
of the metric on the boundary forbids the inclusion of normalizable 
gravity fluctuations. 

The  AdS/CFT correspondence and the Randall Sundrum model can be 
understood as complementary models.
In particular, Duff and  Liu have shown that they share equivalent 
corrections to the Newton's gravitational law\cite{DL}.
A common feature of both AdS/CFT and  Randall Sundrum scenarios 
is that the appropriate 
description of AdS space involves the use of Poincare coordinate system.
This system allows a very  useful definition 
of the bulk/boundary mapping in the AdS/CFT case  
and also  a simple localization for the branes in the Randall Sundrum model.
However, the use of such a coordinate system involves some subtle aspects.
We have considered in a recent letter the quantization of fields in
AdS space using Poincare coordinates \cite{BB} and found the non trivial
result that despite the infinite range of the axial coordinate,
the corresponding spectrum is discrete.
This happens because a consistent quantization in AdS space is only 
possible if one includes a boundary where a vanishing flux of
information from (or to) outside the AdS space 
can be imposed\footnote{We will comment in section 5 on the case of 
D-brane approach to   black p-branes in which case the spectrum 
is continuous and its relation with AdS/CFT correspondence.}. 
In terms of Poincare coordinates, this requires
the introduction of an extra point associated with the axial 
coordinate infinity.
This is only possible by using more than one Poincare coordinate chart.
One has to define the axial coordinate of the first chart stopping 
at some (arbitrary) value, leaving the rest of the space to be mapped 
by other charts.
Here we will discuss this problem further detailing a case where one
can see that the spectrum of eigenfunctions of some space changes 
from discrete to continuous if we map it to coordinate that 
exclude a "point at infinity".
We will also discuss the fact that we have different possible choices
for the quantum fields corresponding to solutions of different boundary 
value problems and not only the particular kind of solution that we 
presented  in \cite{BB}. 

Considering the case of the Randall Sundrum model, the four
dimensional world as we perceive it  corresponds 
to one of the  3-branes whereas the other  acts as a regulating brane. 
So, the slice of AdS space between the branes is compact in the axial 
direction and  bulk fields in this region will have a discrete spectrum,
as discussed by Goldberger and Wise \cite{GW1}. 
These authors  proposed also that such fields with quartic interactions on 
the branes can lead to a mechanism of stabilization of the radius 
$r_c$ of the second brane \cite{GW2}.
The possibility of defining the Randall Sundrum model without
the second brane was discussed in \cite{DK}.
A very recent discussion about bulk fields in the Randall Sundrum scenario
can be found in \cite{Ran}.
In order to gain some understanding on the role of the second brane,
we will discuss here what happens with the field spectrum when we 
take the limit of infinite distance between the two branes taking 
into account the non triviality of the Poincare coordinates.

The article is organized as follows. In section 2 we will review the 
basic properties of AdS and discuss the counting of degrees of freedom
in this space and its relation with the bulk/boundary correspondence.
In section 3 we are going to discuss the mapping of a
compact space into an open space plus a point at infinity, that illustrates
the case of AdS space when represented in Poincare coordinates.  
In section 4 we discuss the quantization of fields in AdS bulk taking
into account the need of multiple charts.
We also discuss how one can choose different boundary conditions
to define complete sets of eigenfunctions and then build up quantum fields.
In section 5 we comment on the difference between the physical setting
of the present model and that of absorption of particles by black p-branes
where a continuous spectrum is found.
In section 6 we will study the implication of our results on the  
spectrum of fields living in the bulk of a Randall-Sundrum
scenario. In particular we will discuss the case when the second brane 
goes to infinity. Some final remarks and conclusions are presented 
in section 7.

%%%%%%%%%%%%%%%%%%%%%%%%%%%%%%%%%%%%%%%%%%%%%%%%%%%%%%%%%%%%%%%%%%%%%%%%%
%%%%%%%%%%%%%%%%%%%%%%%%%%%%%%%%%%%%%%%%%%%%%%%%%%%%%%%%%%%%%%%%%%%%%%%%%

%\newpage
\bigskip

\section{AdS space}

The anti-de Sitter spacetime of $\,n+1$ dimensions is a space of 
constant negative curvature that can be represented 
as the hyperboloid ($\Lambda\,=\,$constant)
\begin{equation}
X_0^2 + X_{n+1}^2 - \sum_{i=1}^n X_i^2\,=\,\Lambda^2
\end{equation}

\noindent embedded in a flat $n+2$ dimensional space with metric
\begin{equation}
ds^2\,=\, - d X_0^2 - dX_{n+1}^2 + \sum_{i=1}^n dX_i^2.
\end{equation}

Two coordinate systems are often used for $AdS_{n+1}$. 
First, the  so called global coordinates $\,\rho,\tau,\Omega_i\,$  
can be defined as \cite{Malda2,Pe}
\begin{eqnarray}
\label{global}
X_0 &=& \Lambda \,\sec\rho\, \cos \tau \nonumber\\
X_i &=& \Lambda \,\tan \rho\, \,\Omega_i\,\,\,\,\,\,\,\,
(\,\sum_{i=1}^n \,\Omega^2_i\,=\,1\,) \nonumber\\
X_{n+1} &=& \Lambda \sec \rho \,\sin\tau \,,
\end{eqnarray}

\noindent with ranges $0\le \rho <\pi/2$ and $0\le\tau< 2\pi\,$.
Quantization of fields in  compactified AdS space (including  
the hypersurface boundary $\rho = \pi/2\,$) 
using these coordinates was discussed in \cite{QAdS1,QAdS2}. 
The coordinate $\tau$, identified with the time coordinate,
 has a finite range in the above prescription.
This is commonly remedied by considering copies of the compact AdS space
glued together along the $\tau $ direction resulting in the 
covering AdS space in which the time coordinate is non compact.
When we refer to AdS space in this article we will actually mean this
covering space.

Second, the Poincar\'e coordinates 
$\,z \,,\,\vec x\,,\,t\,$ can be introduced by
\begin{eqnarray}
\label{Poincare}
X_0 &=& {1\over 2z}\,\Big( \,z^2\,+\,\Lambda^2\,
+\,{\vec x}^2\,-\,t^2\,\Big)
\nonumber\\
X_i &=& {\Lambda x^i \over z}
\nonumber\\
X_n &=& - {1\over 2z}\,
\Big( \,z^2\,-\,\Lambda^2\,+\,{\vec x}^2\,-\,t^2\,\Big)
\nonumber\\
X_{n+1} &=& {\Lambda t \over z}\,,
\end{eqnarray}

\noindent where $\vec x $ with $n-1$ components and  $t$
range from $-\infty $ to $+\infty$ , while $0 \le z < \infty $.
These coordinates are useful for the AdS/CFT correspondence 
\cite{GKP,Wi} and in this case the $\,AdS_{n+1}\,$ measure with 
Lorentzian signature reads\footnote{Another form for Poincare coordinates 
is possible through the mapping $y=\Lambda\ln(z/\Lambda)$, 
which is commonly used in the Randall-Sundrum model, 
as we are going to discuss in section \ref{RS}.}
\begin{equation}
\label{metric}
ds^2=\frac {\Lambda^2 }{( z )^2}\Big( dz^2 \,+(d\vec x)^2\,
- dt^2 \,\Big)\,.
 \end{equation}

\noindent 
Then, in these coordinates, the $AdS$ boundary corresponds to the 
region $z = 0$, described by usual Minkowski 
coordinates $\vec x$ , $t$   plus a ``point'' at infinity 
($z\,\rightarrow\,\infty\,$).
It should  be remarked that the metric is 
not defined at $z = 0\,$ and actually all the calculations in the 
AdS/CFT correspondence are actually done first at some small
$z\,=\,\delta\,$ that in some cases, as when defining boundary 
correlators, is taken to zero after the calculations\cite{MV,FMMR,BB1}

The holographic mapping between the AdS bulk and the corresponding 
boundary \cite{HOL1,HOL2,HOL3}, that are two manifolds of different 
dimensionality, is only possible because the metric is such that 
"volumes" are proportional to "areas". 
We can understand this by counting  the degrees of freedom of the 
bulk volume and those of  the boundary hypersurface. 
In order to compare these quantities that are actually both infinite 
we take a discretized version where the space is not continuous but rather
a discrete array of volume cells.
We take the boundary  at $\,z\,=\,\delta\,$.
There we take a hypersurface of area 
$\,\Delta A \,$ corresponding to variations 
$\,\Delta x^1 \, ... \, \Delta x^{n-1} \,$ in the space coordinates:
\begin{equation}
\,\Delta A \,=
\left(\frac{\Lambda}{\delta}\right)^{n-1}
\,\Delta x^1 \, ... \, \Delta x^{n-1} \,
\end{equation}
 
\noindent 
and calculate the corresponding volume from 
$z = \delta$ to $\infty$, finding
\begin{equation}
\Delta V \,=\,\Lambda \,{\Delta A \over n-1}.
\end{equation}

\noindent This is the expected result that the volume is proportional 
to the area in the bulk / boundary correspondence for a fixed $\Lambda$ 
(see Fig. 1).

%%%%%%%%%%%%%%%%%%%%%%%%%%%%%%%%%%%%%%%%%%%%%%%%%%%%%%%%%%%%
%%%%%%%%%%%%%%%%%%%%%%%%  Figure 1%%%%%%%%%%%%%%%%%%%%%%%%%%
%%%%%%%%%%%%%%%%%%%%%%%%%%%%%%%%%%%%%%%%%%%%%%%%%%%%%%%%%%%%
\
\setlength{\unitlength}{0.06in}
\begin{picture}(50,40)(-20,0)
\label{warpfig}
\rm
\put(22,32){Figure 1}
%%%%%%%%%%%%%%%%%%%  Surface %%%%%%%%%%%%%%%%%%%%%%%%%%%%%%%%
\bezier{1000}(1,25)(-1.5,24)(-2,-1)
\bezier{1000}(-2,-1)(-1.5,-24)(1,-23)
\bezier{1000}(1,-23)(2.5,-24)(3,2)
\bezier{1000}(3,2)(2.5,24)(1,25)
\put(-1.5,5){$\Delta A$}
\put(10,5){$\Delta V$}
%%%%%%%%%%%%%%%%%%%Eixo Horizontal%%%%%%%%%%%%%%%%%%%%%%%%%%%%
\put(0,1){\vector(1,0){54}}
\put(1,0){\line(0,1){2}}
\put(0,-3){$\delta$}
\put(55,-1){$z$}
%%%%%%%%%%%%%%%%%%%%Dados%%%%%%%%%%%%%%%%%%%%%%%%%%%%%%%%%%%%%%
\bezier{800}(1,25)(5,5)(50,2)
\bezier{800}(1,-23)(5,-3)(50,0)
\end{picture}
\vskip 4cm
\noindent Fig. 1:
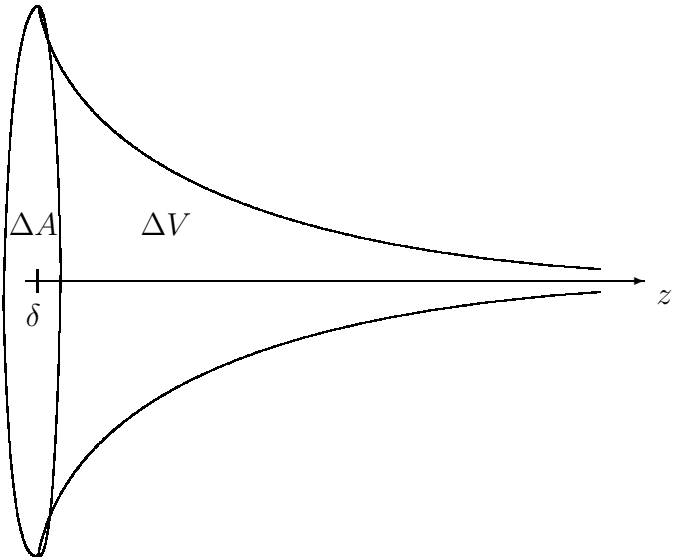
{\it The Area $\Delta A$ on the hypersurface at $z=\delta$ and
the corresponding AdS volume $\Delta V$.}
%%%%%%%%%%%%%%%%%%%%%%%%%%%%%%%%%%%%%%%%%%%%%%%%%%%%%%%%%%%%%%%%
\vskip 1cm
%%%%%%%%%%%%%%%%%%%%%%% end of figure 1 %%%%%%%%%%%%%%%%%%%%%%%%
%%%%%%%%%%%%%%%%%%%%%%%%%%%%%%%%%%%%%%%%%%%%%%%%%%%%%%%%%%%%%%%%

Now we can count the degrees of freedom by splitting 
$\Delta V$ in $\ell$ pieces of equal volume corresponding to cells whose 
boundaries are hypersurfaces located at
\begin{equation}
z_j\,=\, {\delta\over \sqrt[n-1]{1- j/\ell}}\,\,\,\,,
\end{equation}

\noindent with $\,j\,=\,0, 1,..,\ell -1\,$. Note that the last cell 
extends to infinity. 
These volume cells can be mapped into the area $\,\Delta A\,$ 
by also dividing it in $\ell$ parts. We can take $\ell$ to be arbitrarily 
large.  This way  we find a one to one
mapping between degrees of freedom of bulk and boundary (see Fig. 2).  
This analysis shows why it is possible to holographically map the
degrees of freedom of the bulk on the boundary despite the fact 
that the variable $z$ has an infinite range.
One could then think of the AdS space as corresponding to a "box" 
in terms of degrees of freedom with respect to $z$. By this we mean that
effectively the coordinate $z$ behaves as if it had a finite range.
Then the field spectrum associated with this coordinate
should be discrete contrarily to ones expectation for a variable of 
infinite range.

%%%%%%%%%%%%%%%%%%%%%%%%%%%%%%%%%%%%%%%%%%%%%%%%%%%%%%%%%%%%%%%%%%
%%%%%%%%%%%%%%%%%%%%%%%%  Figure 2 %%%%%%%%%%%%%%%%%%%%%%%%%%%%%%%
%%%%%%%%%%%%%%%%%%%%%%%%%%%%%%%%%%%%%%%%%%%%%%%%%%%%%%%%%%%%%%%%%%
\
\setlength{\unitlength}{0.08in}
\begin{picture}(50,40)(-10,0)
\rm
\put(22,32){Figure 2}
%%%%%%%%%%%%%%%%%%%Eixo Vertical%%%%%%%%%%%%%%%%%%%%
\put(0,-1){\vector(0,1){30}}
\put(1,28){Boundary}
\put(-3,0){0}
%%%%%%%%%%%%%%%%%%%Eixo Horizontal%%%%%%%%%%%%%%%%%%%
\put(-1,1){\vector(1,0){58}}
\multiput(3,0)(0,2.05){10}{\line(0,1){1}}
\put(2,-2){$z_0$}
\multiput(11,0)(0,2.1){6}{\line(0,1){1}}
\put(10,-2){$z_1$}
\multiput(23,0)(0,1.7){4}{\line(0,1){1}}
\put(22,-2){$z_2$}
\multiput(45,0)(0,1.5){2}{\line(0,1){1}}
\put(44,-2){$z_3$}
\put(55,-1){$z$}
%%%%%%%%%%%%%%%%%%%%Dados%%%%%%%%%%%%%%%%%%%%%%%%%%%%
\bezier{1400}(1,25)(5,5)(50,2)
\end{picture}
\vskip 1cm
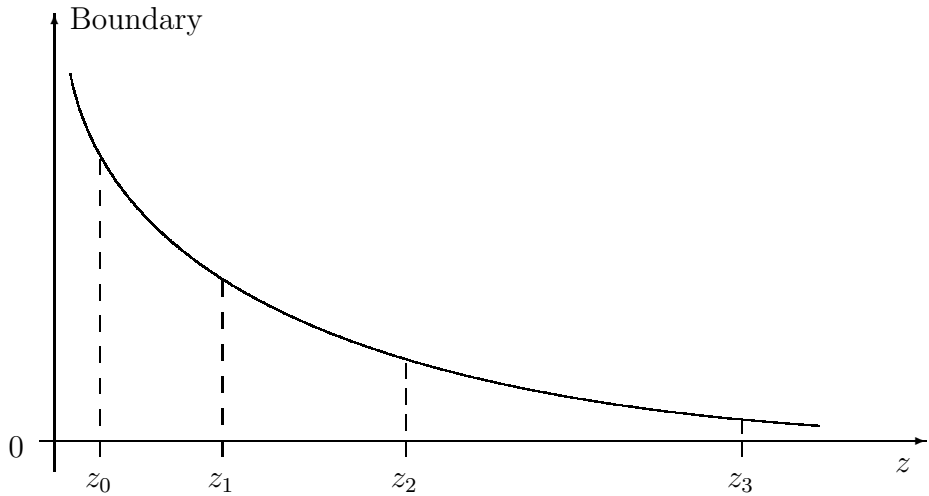
\centerline{Fig. 2: \it z section of equal AdS volume cells.}
%%%%%%%%%%%%%%%%%%%%%%%%%%%%%%%%%%%%%%%%%%%%%%%%%%%%%%%%%%%%%%%%
%%%%%%%%%%%%%%%%%%%% end of figure 2 %%%%%%%%%%%%%%%%%%%%%%%%%%%
%%%%%%%%%%%%%%%%%%%%%%%%%%%%%%%%%%%%%%%%%%%%%%%%%%%%%%%%%%%%%%%%

\bigskip

In the global coordinate context, where the coordinates have  
finite ranges it was shown that a consistent quantization 
can only be obtained if  one  supplements AdS space 
with a boundary ("the wall of the box") in order to impose vanishing
flux of particles and  information there\cite{QAdS1,QAdS2}.
Otherwise massless particles would be able to go to (or come from)  
spatial infinity in finite times and it would thus be impossible 
to define a Cauchy surface. 
This boundary is the hypersurface $\rho\,=\,\pi/2$ 
that corresponds in Poincare coordinates  to the surface $z\,=\,0\,$ 
plus a point defined by the limit $\,z\,\rightarrow\,\infty\,$.
Thus a consistent quantization would require the inclusion of
this extra point at infinity where the flux should be required to vanish.
This is not possible just using one set of Poincar\'e coordinates
because the mapping between these two coordinate systems is not one to one.
The non trivial  topology of AdS space and problems related to using 
coordinates like Poincare are also discussed in \cite{McI}.
In order to understand more precisely this 
problem  and its consequences for the field spectrum we will 
consider in the next chapter a simpler illustrative problem.

\section{Mapping of a compact space into an open space plus 
a point at infinity}
\label{mapping}

If we start with a description of compactified AdS space 
(including the hypersurface $\rho\,=\,\pi/2$) 
in terms of global coordinates and then transform
to Poincare coordinates we will be  considering  a mapping that is not 
one to one. 
In order to have a clear understanding of this point and to 
illustrate its consequences for the field spectrum we will 
first consider a simpler example where
a similar  situation happens: the case of a 
one dimensional compact space where  a change of coordinates
apparently turns the spectrum of eigenfunctions from 
discrete to continuous.
We will see that this happens precisely when we map 
the compact space into an open set of infinite range plus  a  point 
at infinity. At least, two coordinate charts will be necessary in order 
to find a one to one mapping and thus a correct description of the 
associated functional space.

Let us consider the compact interval 
$S\,=\,\{ x\, \vert \,\, 0 \,\le \,x \,\le L\,\}$ 
along the cartesian coordinate $x$.
Considering the class of functions $f$ that are not 
singular in $S$ we know that a basis can be formed 
 by solving a boundary value problem for some 
 Sturm Liouville operator on $S$, corresponding in general to 
some linear combination of the functions and their derivatives 
vanishing 
at $x\,=\,0 $ and $x\,=\, L\,$. Any function $f $ 
can be represented in the open set $S -\{0\} 
-\{L\}$ as, for example:
\begin{equation}
f(x)\,=\,\sum_n  a_n \sin ({n\pi x\over L}) 
\end{equation}

\noindent or, alternatively, changing the boundary condition
\begin{equation}
f(x)\,=\,\sum_n  b_n \cos ({n\pi x\over L}) 
\end{equation}

\noindent or some sinusoidal function corresponding to a mixed 
boundary condition. The point is that, given the compact interval $S$,
an arbitrary non-singular function  in $S$ can be expanded, 
except possibly at the
end points, as a discrete series of eigenfunctions.
This in some sense defines the "dimension" of this functional space.

Let us however see what happens if we  introduce the variable 
\begin{equation}
\label{map1}
u \,= {1\over L - x }
\end{equation}

\noindent and map the interval $S$ in the interval 
$S^\prime = \{ u\, \vert 1/L \le u < \infty\,\}$. This map induces a 
metric in $S^\prime$ (considering the original set $S$ as of unity 
measure).
\begin{equation}
\label{metric1}
ds^2 \,=\, dx^2\,=\,{{du}^2\over u^4}
\end{equation}

\noindent 
The point $x = L$ is "mapped" to $u \,\rightarrow \,\infty\,$.
Now if we naively consider the spectrum of eigenfunctions 
by just looking at $S^\prime$ we would
conclude that it is continuous because the interval is not compact.
An interesting way of understanding what is happening is just to try 
reversing the mapping between the two sets.  
If we want to  naively reverse the mapping from $S^\prime$ to $S$
by just looking at eq. (\ref{map1}) we would not get the point 
$x\,=\,L\,$ back.
We would then  find an open (at one side) set, corresponding to a 
continuous spectrum of eigenfunctions. This difference in the 
spectrum would be a 
consequence of the absence of point $x = L $ and thus the lack of 
a condition of non-singularity there and thus an increase in the
set of admissible functions. So we would be changing the 
functional space by naively representing the set by the 
coordinate $u$ because the mapping of 
$S$ on $S^\prime\,$ is not one to one. 
It is interesting to observe also that the point $x = L $ is mapped
into a limit of singular (vanishing) metric in the $u$ coordinate.
Even in the case that such a singular point could be included in the 
coordinate system, a second chart should be introduced\cite{HE}.

So it is not possible to map the whole interval $S$ into $S^\prime$ with 
just one chart of the coordinate $u$  but rather we should consider
at least two charts.
We can, for example define the range of $u$ to be:
$ {1\over L} \le u \le {1\over L - R}$ that would map 
the set $0\,\le x \,\le R\,$. Then we can map the rest of
$S$ with another variable: $\,v\,=\,1/x$ with range 
$ 1/L \,\le\, v\, \le\, 1/R $ and induced metric 
\begin{equation}
ds^2 \,=\, dx^2 \,=\, {{dv}^2\over v^4}.
\end{equation}

Now the mapping
\begin{eqnarray}
x \mapsto \left\{ 
\begin{array}{ll}
u = 1/(L - x) & 0\le x \le R \\
v = 1/ x 	  & R \le x \le L
\end{array}
\right.
\end{eqnarray}

\noindent 
is one to one and in both charts we can find discrete
basis of eigenfunctions. 
We can take $R$  arbitrarily close to $L$ but not equal.
This mapping does really reproduces the interval $S$ and indeed gives the 
same kind of spectrum of eigenfunctions. In other words they share 
to the same functional space. 

The first tentative mapping was not one to one and we were 
loosing the possibility of imposing any kind of boundary condition at 
$x\,=\,L\,$, and thus considering a different (larger) functional space.
This point plays the role of a point at infinity for the coordinate
$\,u\,$ and it is interesting to observe that removing one point we 
indeed get a larger functional space. This fact will be very important 
when we consider the AdS case where the functional space 
has to be such that a (one to one) mapping with the functional space 
of the boundary should be possible.
A representation with just one Poincare chart would, in the same 
way as in the present example, not be appropriate as it would actually
lead to  a larger functional space than that of the compactified AdS.

Also important to remark, for understanding the AdS situation,
is the fact that we are not saying that we must include some cut off 
in the interval $S$ itself. We just have to use at least two coordinate 
charts. So the distance $R$ has no significance for the set $S$ although
some value should be chosen in order to be possible to find a one to one 
representation in terms of the coordinates  $u$ and $v$. 

For completeness we observe also that if our original interval did 
not include the point $x = L$ the mapping with $S^\prime $ would be 
one to one and  there is no need for two charts.
Even if we decide to do so, the second chart would be an open set
and thus we would conclude that the spectrum would be 
continuous as expected for an original open set that really would 
correspond to a larger functional space.

%%%%%%%%%%%%%%%%%%%%%%%%%%%%%%%%%%%%%%%%%%%%%%%%%%%%%%%%%%%%%%%%%%%%%
%%%%%%%%%%%%%%%%%%%%%%%%%%%%%%%%%%%%%%%%%%%%%%%%%%%%%%%%%%%%%%%%%%%%%
 
\bigskip

\section{Quantum fields in the AdS space}

\bigskip
 
Now we turn to the question of quantizing fields in the AdS bulk.
This problem was originally investigated in \cite{QAdS1,QAdS2}
using the global coordinates, eq. (\ref{global}).
The main problem that one finds when quantizing fields in AdS is that
this space does not admit a Cauchy surface and consequently suffers from
the Cauchy problem \cite{HE}. This problem  was  circumvented by
compactifying the AdS space including the hypersurface $\rho=\pi/2$.
With this prescription it is possible to show that energy and information
are conserved and a consistent quantization is possible.
The problem we want to discuss here is the quantization of fields in AdS
bulk described by Poincare coordinates, eq. (\ref{Poincare}).
If the mapping between compactified AdS in global and Poincare coordinates
were one to one this problem would be trivial. However, as discussed in
the previous sections, these coordinates have infinite
range and  the infinite limit of the axial coordinate 
$(z\to\infty)$ corresponds to a point in the AdS boundary that is not 
properly represented by the chart given by eq. (\ref{Poincare}). 
So using the discussion of section {\bf \ref{mapping}} on the mapping 
of a compact
space (here the compactified AdS space in global coordinates) 
into an open set plus a point
at infinity (here the AdS in Poincare coordinates) we conclude that 
a consistent quantization in the whole 
AdS space must involve more than one Poincare chart in order to preserve
the dimension of the functional space. 

Considering the interval $\delta \le z < \infty$, 
we find a way of introducing another chart, with the help of 
the example of the previous section, by first introducing an auxiliary 
coordinate $\,\alpha \,$ as

\begin{equation}
\label{z}
z \,=\, {1\over {\textstyle 1\over\textstyle\delta} \,-\,\alpha}\,\,.
\end{equation}

\noindent with $ 0 \le \alpha \le  1/\delta \,$.

In order to have a proper definition
of the point at infinity $z\to\infty$, we can introduce a new chart as 
\begin{equation}
z^\prime \,=\, {1\over \alpha }\,.
\end{equation}

Now the Poincare charts correspond to the system
of eqs. (\ref{Poincare}), (\ref{metric}) plus a second one where 
$z$ is rewritten in terms of $z^\prime$ from
\begin{eqnarray}
{1\over z^\prime} = {1\over \delta} - {1\over z} 
\end{eqnarray}

\noindent with ranges  $\delta \le z \le R \,$ 
and $\delta \le z^\prime \le R^\prime \,$, where 
$R^\prime= \delta R /( R - \delta )\,$ (see Fig. 3).

We can take $R$ arbitrarily large and map as much of AdS space as we want 
into just one chart but the fact that $R$ is finite implies the 
discretization of the spectrum associated with the coordinate $z$.
This reduces the dimensionality of the functional space.
A similar problem was discussed by Gell-Mann and Zwiebach\cite{GZ}
in the context of dimensional reduction induced by a sigma model.

Note that the region $0\le z \le \delta$ is not covered by the above charts.
In order to describe this region with the auxiliary variable $\alpha$ we can
still use eq. (\ref{z}), but with $-\infty < \alpha \le 0$ (see Fig. 4). 
However in this case a non compact region $\alpha$ is mapped into a compact 
$z$ region, so that the mapping is not one to one. If this is required one 
has to define another $\alpha$ chart.

%%%%%%%%%%%%%%%%%%%%%%%%%%%%%%%%%%%%%%%%%%%%%%%%%%%%%%%%%%%%%%%%%%%%%%%%%%%%
%%%%%%%%%%%%%%%%%%%%%%%%%%       FIGURE 3     %%%%%%%%%%%%%%%%%%%%%%%%%%%%%%
%%%%%%%%%%%%%%%%%%%%%%%%%%%%%%%%%%%%%%%%%%%%%%%%%%%%%%%%%%%%%%%%%%%%%%%%%%%%
\
\setlength{\unitlength}{0.08in}
\vskip 1.5cm
\begin{picture}(0,0)(-2,0)
\rm
\put(32,7){Figure 3}
%%%%%%%%%%%%%%%%%%%%%%%%%%%%%%% A %%%%%%%%%%%%%%%%%%%%%%%%
%%%%%%%%%%%%%%%%%%%      Eixo Horizontal   Z   %%%%%%%%%%%%%%
\put(-5,1){a)}
\put(-1,1){\vector(1,0){18}}
\put(0,0.9){\line(1,0){16.5}}
\put(0,1.05){\line(1,0){16.5}}
\put(0,0.5){\line(0,1){1}}
\put(-0.5,-2){$\delta$}
\put(11,0.5){\line(0,1){1}}
\put(10,-2){$R$}
\put(19,.5){$z$} 
%%%%%%%%%%%%%%%%%%%%%%     alpha     %%%%%%%%%%%%%%%%%%%%%%%%%
\put(29,1){\vector(1,0){14}}
\put(30,0.9){\line(1,0){10}}
\put(30,1.05){\line(1,0){10}}
\put(30,0.5){\line(0,1){1}}
\put(29,-2){$0$}
\put(40,0.5){\line(0,1){1}}
\put(38.5,-2){$1/\delta$}
\put(44,.5){$\alpha$} 
%%%%%%%%%%%%%      Eixo Horizontal   Z(prime)   %%%%%%%%%%%%%%
\put(73,1){\vector(-1,0){18}}
\put(71,.9){\line(-1,0){15.5}}
\put(71,1.05){\line(-1,0){15.5}}
\put(53,.5){$z^\prime$} 
\put(62,0.5){\line(0,1){1}}
\put(61,-2){$R^\prime$}
\put(71,0.5){\line(0,1){1}}
\put(70,-2){$\delta$}
\end{picture}
\vskip 1cm
%%%%%%%%%%%%%%%%%%%%%%%%%%%%%%%%%%%%%%%%%%%%%%%%%%%%%%%%%%%%%%%%%
\begin{picture}(0,0)(-2,0)
\rm
%%%%%%%%%%%%%%%%%%%%%%%%%%%%%%% B %%%%%%%%%%%%%%%%%%%%%%%%
%%%%%%%%%%%%%%%%%%%      Eixo Horizontal   Z   %%%%%%%%%%%%%%
\put(-5,1){b)}
\put(-1,1){\vector(1,0){18}}
\thicklines
\put(0,0.95){\line(1,0){11}}
\put(0,1.05){\line(1,0){11}}
\thinlines
\put(0,0.5){\line(0,1){1}}
\put(-0.5,-2){$\delta$}
\put(11,0.5){\line(0,1){1}}
\put(10,-2){$R$}
\put(19,.5){$z$} 
%%%%%%%%%%%%%%%%%%%%%%   relation    %%%%%%%%%%%%%%%%%%%%%%%%%
\put(22,0.5){$\Longleftrightarrow$}
%%%%%%%%%%%%%%%%%%%%%%     alpha     %%%%%%%%%%%%%%%%%%%%%%%%%
\put(29,1){\vector(1,0){14}}
\thicklines
\put(30,0.9){\line(1,0){6}}
\put(30,1.05){\line(1,0){6}}
\thinlines
\put(30,0.5){\line(0,1){1}}
\put(29,-2){$0$}
\put(36,0.5){\line(0,1){1}}
\put(35,-2){$1\over R^\prime$}
\put(40,0.5){\line(0,1){1}}
\put(39.5,-2){$1\over\delta$}
\put(44,.5){$\alpha$} 
%%%%%%%%%%%%%      Eixo Horizontal   Z(prime)   %%%%%%%%%%%%%%
\put(73,1){\vector(-1,0){18}}
\thicklines
\put(62,.9){\line(-1,0){6.5}}
\put(62,1.05){\line(-1,0){6.5}}
\thinlines
\put(53,.5){$z^\prime$} 
\put(62,0.5){\line(0,1){1}}
\put(61,-2){$R^\prime$}
\put(71,0.5){\line(0,1){1}}
\put(70,-2){$\delta$}
\end{picture}
\vskip 1cm
%%%%%%%%%%%%%%%%%%%%%%%%%%%%%%%%%%%%%%%%%%%%%%%%%%%%%%%%%%%%%%%%%%
%%%%%%%%%%%%%%%%%%%%%%%%%%%%%%%%%%%%%%%%%%%%%%%%%%%%%%%%%%%%%%%%%%
\begin{picture}(0,0)(-2,0)
\rm
%%%%%%%%%%%%%%%%%%%%%%%%%%%%%%% C %%%%%%%%%%%%%%%%%%%%%%%%
%%%%%%%%%%%%%%%%%%%      Eixo Horizontal   Z   %%%%%%%%%%%%%%
\put(-5,1){c)}
\put(-1,1){\vector(1,0){18}}
\put(11,0.9){\line(1,0){5.5}}
\put(11,1.05){\line(1,0){5.5}}
\put(0,0.5){\line(0,1){1}}
\put(-0.5,-2){$\delta$}
\put(11,0.5){\line(0,1){1}}
\put(10,-2){$R$}
\put(19,.5){$z$} 
%%%%%%%%%%%%%%%%%%%%%%     alpha     %%%%%%%%%%%%%%%%%%%%%%%%%
\put(29,1){\vector(1,0){14}}
\thicklines
\put(36,0.9){\line(1,0){4}}
\put(36,1.05){\line(1,0){4}}
\thinlines
\put(30,0.5){\line(0,1){1}}
\put(29,-2){$0$}
\put(36,0.5){\line(0,1){1}}
\put(35,-2){$1\over R^\prime$}
\put(40,0.5){\line(0,1){1}}
\put(39.5,-2){$1\over\delta$}
\put(44,.5){$\alpha$} 
%%%%%%%%%%%%%%%%%%%%%%   relation    %%%%%%%%%%%%%%%%%%%%%%%%%
\put(47,0.5){$\Longleftrightarrow$}
%%%%%%%%%%%%%      Eixo Horizontal   Z(prime)   %%%%%%%%%%%%%%
\put(73,1){\vector(-1,0){18}}
\thicklines
\put(71,.9){\line(-1,0){9}}
\put(71,1.05){\line(-1,0){9}}
\thinlines
\put(53,.5){$z^\prime$} 
\put(62,0.5){\line(0,1){1}}
\put(61,-2){$R^\prime$}
\put(71,0.5){\line(0,1){1}}
\put(70,-2){$\delta$}
%%%%%%%%%%%%%%%%%%%%%%%%%%%%%%%%%%%%%%%%%%%%%%%%%%%%%%%%%%%%%%%%%
%%%%%%%%%%%%%%%%%%%%%%%%%%%%%%%%%%%%%%%%%%%%%%%%%%%%%%%%%%%%%%%%%
\end{picture}
\vskip 1cm
%%%%%%%%%%%%%%%%%%%%%%%%%%%%%%%%%%%%%%%%%%%%%%%%%%%%%%%%%%%%%%%%%%%%%%%%%%%%
%%%%%%%%%%%%%%%%%%%%%%%%%%% FIGURE 3 CAPTION  %%%%%%%%%%%%%%%%%%%%%%%%%%%%%%
\noindent 
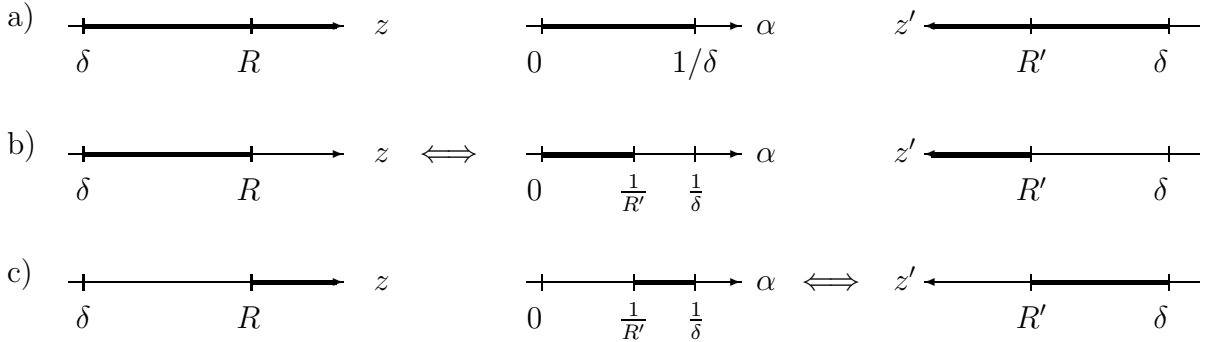
Fig. 3: {\it The corresponding intervals on $z$, $\alpha$ and $z^\prime$. 
a) The mappings are not one to one, since $z$ and $z^\prime$ are non compact 
while $\alpha$ is compact.   
b) The compact region $\delta \le z \le R$ has a one to one mapping 
(indicated by the double arrow) with $0 \le \alpha\le 1/R^\prime$,
corresponding to one Poincare chart.
c) $z^\prime$ in the region $\delta \le z^\prime \le R^\prime$ has a one to
 one mapping with $1/R^\prime \le \alpha \le 1/\delta$, corresponding to a 
second Poincare chart.}
\vskip 1cm 
%%%%%%%%%%%%%%%%%%%%%%%%%%%%%%%%%%%%%%%%%%%%%%%%%%%%%%%%%%%%%%%%%%%%%%%%%%%%
%%%%%%%%%%%%%%%%%%%%%%%%%%%%% END OF FIGURE 3 %%%%%%%%%%%%%%%%%%%%%%%%%%%%%%
%%%%%%%%%%%%%%%%%%%%%%%%%%%%%%%%%%%%%%%%%%%%%%%%%%%%%%%%%%%%%%%%%%%%%%%%%%%%

%%%%%%%%%%%%%%%%%%%%%%%%%%%%%%%%%%%%%%%%%%%%%%%%%%%%%%%%%%%%%%%%%%%%%%%%%%%
%%%%%%%%%%%%%%%%%%%%%%%%%%       FIGURE 4     %%%%%%%%%%%%%%%%%%%%%%%%%%%%%
%%%%%%%%%%%%%%%%%%%%%%%%%%%%%%%%%%%%%%%%%%%%%%%%%%%%%%%%%%%%%%%%%%%%%%%%%%%
\
\setlength{\unitlength}{0.08in}
\vskip 1.5cm
\begin{picture}(0,0)(-2,0)
\rm
\put(32,7){Figure 4}
%%%%%%%%%%%%%%%%%%%%%%%%%%%%%%% A %%%%%%%%%%%%%%%%%%%%%%%%
%%%%%%%%%%%%%%%%%%%      Eixo Horizontal   Z   %%%%%%%%%%%%%%
\put(-1,1){\vector(1,0){18}}
\put(0,0.9){\line(1,0){11}}
\put(0,1.05){\line(1,0){11}}
\put(0,0.5){\line(0,1){1}}
\put(-0.5,-2){0}
\put(11,0.5){\line(0,1){1}}
\put(10.5,-2){$\delta$}
\put(19,.5){$z$} 
%%%%%%%%%%%%%%%%%%%%%%     alpha     %%%%%%%%%%%%%%%%%%%%%%%%%
\put(29,1){\vector(1,0){14}}
\put(29,0.9){\line(1,0){11}}
\put(29,1.05){\line(1,0){11}}
\put(40,0.5){\line(0,1){1}}
\put(39.5,-2){0}
\put(44,.5){$\alpha$} 
%%%%%%%%%%%%%      Eixo Horizontal   Z(prime)   %%%%%%%%%%%%%%
\put(73,1){\vector(-1,0){18}}
\put(62,.9){\line(1,0){11}}
\put(62,1.05){\line(1,0){11}}
\put(53,.5){$z^\prime$} 
\put(62,0.5){\line(0,1){1}}
\put(61.5,-2){0}
\end{picture}
\vskip 1cm
%%%%%%%%%%%%%%%%%%%%%%%%%%%%%%%%%%%%%%%%%%%%%%%%%%%%%%%%%%%%%%%%%%%%%%%%%
%%%%%%%%%%%%%%%%%%%%%%%%%%%%%%%%%%%%%%%%%%%%%%%%%%%%%%%%%%%%%%%%%%%%%%%%%%
%%%%%%%%%%%%%%%%%%%%%%%%%%% FIGURE 4 CAPTION  %%%%%%%%%%%%%%%%%%%%%%%%%%%%
\noindent 
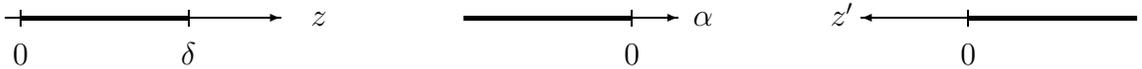
\centerline{Fig. 4: \it Corresponding intervals for  $0\le z\le \delta$, 
into $\alpha$ and $z^\prime$.}
\vskip 1cm 
%%%%%%%%%%%%%%%%%%%%%%%%%%%%%%%%%%%%%%%%%%%%%%%%%%%%%%%%%%%%%%%%%%%%%%%%%%
%%%%%%%%%%%%%%%%%%%%%%%%%%%%% END OF FIGURE 4 %%%%%%%%%%%%%%%%%%%%%%%%%%%%
%%%%%%%%%%%%%%%%%%%%%%%%%%%%%%%%%%%%%%%%%%%%%%%%%%%%%%%%%%%%%%%%%%%%%%%%%%

In \cite{BB}  we discussed the case of quantum scalar fields in 
anti de Sitter space in Poincare coordinates but considering 
a chart where the axial coordinate has the range  $0\le z \le R$ 
and choosing as a functional basis for the fields a particular 
boundary value problem corresponding to
functions vanishing at $ z = 0 $  and $ z = R \,$. 
We will study here different boundary conditions on an interval
 $\delta \le z \le R$. This way we avoid the surface
$ z = 0\, $ where the metric is not defined. Also by an appropriate
 choice of $\delta $ and boundary conditions we  
reproduce the Randall Sundrum scenario. 

Let us  consider a massive scalar field $\phi$ in the $\,AdS_{n+1}\,$
spacetime described by Poincar\'e coordinates with action
\begin{equation}
\label{action1}
I[\phi ]\,=\, {1\over 2} \int d^{n+1}x \sqrt{g}\,
\left(\partial_\mu \phi \, \partial^\mu \phi
+m^2\,\phi^2 \right)
\,\,,
\end{equation}
  
\noindent where we take $x^0\,\equiv\,z\,,\,x^{n}\,\equiv\,t\,$,
$\sqrt{g}\,=\,(x^0)^{-n-1}\,$ and $\mu\,=\,0,1,...,n\,$.
The classical equation of motion then reads                                                                                                                                                                                                                                                                                                                                                                                                          
\begin{equation}
\label{motion}
\left(\nabla_\mu \nabla^\mu - m^2\right) \phi\,
=\,{1\over \sqrt{g}} 
\partial_\mu 
\Big( \, \sqrt{g} \partial^\mu \phi \,\Big) 
- m^2\phi\,=\,0\,\,
\end{equation}

\noindent and one finds solutions\cite{BKL,BKLT} for the interval
$\delta \le z \le R$ as a linear combination ($c$ and $d$ are arbitrary
constants)

\begin{equation}
\label{Sol1}
\Phi(z, \vec x , t) =\, e^{-i\omega t\,+\,i\vec k \cdot \vec x} z^{n/2} 
\,\Big( \, c J_{\nu}(u z) \,+\, d Y_{\nu}(u z) \,\Big)\,,
\end{equation}

\noindent where $J_{\nu}(u z)$ and $Y_{\nu}(u z)$ are, respectively, 
the Bessel and Neumann functions of order $\nu=\frac 12\sqrt{n^2+4m^2}$ and 
$u\,\equiv \,\sqrt{ \omega^2\,-\,{\vec k}^2\,}\,$ with 
$\omega^2\,>{\vec k}^2$.

In the region considered: $\delta\,\le\,z\,\le R\,$ we can find a functional 
basis for representing fields by integrating over $\vec k $ the 
solutions of any boundary value problem in the coordinate $z$  of the 
general form

\begin{eqnarray}
\label{BC1}
A \, \Phi (z= \delta , \vec x , t) &+&
B \, {{\partial \Phi\over \partial z} (z, \vec x , t)}\vert_{z = \delta}
\,=\, F(\vec x, t)
\\
\label{BC2}
C \, \Phi (z= R , \vec x , t) &+&
D \, {{\partial \Phi\over \partial z} (z, \vec x , t)}\vert_{z = R}\,=\,
 G(\vec x, t)\,\,\,.
\end{eqnarray}

\noindent Here in this work we will choose for simplicity  $F(\vec x, t)= 
G(\vec x, t)=0$. 
Any non trivial choice of the constants $A$, $B$, $C$, $D$
leads to a basis of eigenfunctions for the  
operator appearing in the equation of motion (\ref{motion}).
If we choose, for example, $B\,=\,D\,=\,0\,$ we find the condition

\begin{equation}
\label{Sol2}
 J_{\nu}(u_p \delta) \, Y_{\nu}(u_p R ) \,-
\, J_{\nu}(u_p R ) \, Y_{\nu}(u_p \delta ) \,=\,0\,.
\end{equation}

\noindent The roots of this equation define the possible values of $u_p$,
restricting the spectrum of the field. Choosing these eigenfunctions, we
can write the quantum fields as

\begin{eqnarray}
\label{QF}
\Phi(z,\vec x,t) &=&  
\sum_{p=1}^\infty \,
\int {d^{n-1}k \over (2\pi)^{n-1}}\,
 z^{n/2} N (\vec k , u_p )\Big[ J_\nu (u_p z ) \,- \, 
\Big( { J_\nu (u_p \delta )\over
   Y_{\nu}(u_p \delta )} \Big)\,  Y_{\nu}(u_p z ) \Big]
\nonumber\\
&&\qquad 
\times \; \lbrace { a_p(\vec k )\ } e^{-iw_p(\vec k ) t +i\vec k \cdot 
\vec x}\,
\,+\,\,{ a_p^\dagger(\vec k )\ } e^{iw_p(\vec k ) t 
-i\vec k \cdot \vec x}\, \rbrace
\end{eqnarray}

\noindent where $w_p(\vec k ) \,=\,\sqrt{ u_p^2\,+\,{\vec k}^2}$ and 
$N(\vec k , u_p )$ is a normalization constant.

Imposing that the operators $a_p(\vec k )\,,\,
a^\dagger_{p^\prime}({\vec k}^\prime)$ 
satisfy the commutation relations
\begin{eqnarray}
\Big[ a_p(\vec k )\,,\,a^\dagger_{p^\prime}({\vec k}^\prime  )
\Big]&=& 2\, (2\pi)^{n-1} w_p(\vec k )   
\delta_{p\,  p^\prime}\,\delta^{n-1} (\vec k -
{\vec k}^\prime )  \nonumber\\
\Big[ a_p(\vec k )\,,\,a_{p^\prime}({\vec k}^\prime  )
\Big] &=& \Big[ a^\dagger_p(\vec k )\,,\,
a^\dagger_{p^\prime}({\vec k}^\prime  ) \Big]\,=\,0
\end{eqnarray}

\noindent we find, for example, for the equal time commutator of the field 
and its time derivative
\begin{equation}
\Big[ \Phi (z,\vec x ,t)\,,\,{\partial\Phi\over \partial t}(z^\prime,
\vec x^\prime ,t)\,\Big]\,=\,i z^{n-1} \delta (z - z^\prime) 
\delta (\vec x - {\vec x}^\prime )\,.
\end{equation}

Other boundary conditions corresponding to different choices of 
the constants in eqs. (\ref{BC1}),(\ref{BC2}) can be chosen and 
would  lead to different  basis for representing the field $\Phi$. 
For  example, the case $A\,=\,C\,= 0$  is connected with the 
Randall-Sundrum model as we are going to discuss in the section 
{\bf 6}. Before continuing this discussion let us make some comments
on the difference between the situation described above, which implies
a discrete spectrum and that of black p-branes, also related to
AdS/CFT correspondence, but where particles have a continuous spectrum.

\section{AdS, Black p-Branes and boundary conditions}

\subsection{AdS and Black p-Branes }

The study of the D-brane formulation of black p-branes was a source
of motivation for the discovery of the AdS/CFT correspondence.
The comparison of the absorption cross section of infalling particles
on D3 branes with perturbative calculations on the brane world 
volume\cite{Kle1}  indicated that  Greens functions of 
Yang Mills theory (with extended supersymmetry) could be calculated 
from  supergravity (see  \cite{Malda2},\cite{Kle2} for a review 
and a wide list of references).

The extremal 3-brane metric in $D=10$ dimensions\cite{HoSt91} may be 
written as

\begin{equation}
ds^2=\left(1+\frac{R^4}{r^4}\right)^{-1/2}(-dt^2+d\vec{x}^2)
+\left(1+\frac{R^4}{r^4}\right)^{1/2}(dr^2 +r^2d\Omega_5^2)
\end{equation}

\noindent where $0\le r <\infty$  with a horizon at $r=0$. 
In the near-horizon region the metric takes the simpler form

\begin{equation}
ds^2=\frac{R^2}{z^2}(-dt^2+d\vec{x}^2+dz^2) + R^2d\Omega_5^2
\end{equation}

\noindent which is an $AdS_5 \times S_5$ geometry, where $z=R^2/r$.

In this study of absorption cross section of particles
on D3 branes of \cite{Kle1} a transmitted wave at 
$z\rightarrow \infty\,$ is associated with particles falling into 
the brane. The flux of particles at infinity is related to the  probability of 
absorption by the horizon. 
It is important to realise the difference between this physical picture 
and the one considered here.
This way we will understand the difference of the field spectrum in the 
two cases.

\subsection{Boundary conditions}

Let us see what is precisely the physical picture that we are considering here
and what is the physical reason that leads to a discrete field spectrum.
We are considering, like the aproach of \cite{Wi},
a realization of AdS/CFT correspondence  defined in just 
a purely AdS space (or more precisely, its covering space).
That means:  there is nothing to absorb particles, or energy, at infinity.
The definition of a consistent quantum field theory in AdS space
requires adding a boundary at  infinity where  one imposes a
vanishing flux of energy.  This is a necessary condition 
in order to have a well posed Cauchy problem with a unique solution\cite{QAdS1,QAdS2}.
Thus one actually needs a compactified
version of the space, as we discussed  in section 2.

In a simple way, we can think that considering just the AdS space, as we are
doing here,  there should be  nothing to absorb particles, or energy 
at spatial infinity. Thus, considering that massless particles would go to spatial
infinity in finite times, one needs to incorporate the idea of \cite{QAdS1,QAdS2} 
of "closing" (compactifying) the AdS space in the Poincare coordinate framework.
The spectrum associated with a compactified coordinate clearly has to be discrete, but 
the way one realizes this in the case of the axial Poincare coordinate $z$ 
is subtle. A compactification of this coordinate is not possible by using 
just one Poincare coordinate chart because one needs to add an extra point at infinity.
As discussed in section 4 we need to stop the coordinate $z$ at some value $R$ and 
map the rest of the space in a second coordinate chart.
It is important to stress that we are not imposing any special
kind of boundary condition at $z=R$. However the fact that we have to cut the coordinate
$z$ at some finite value (or equivalently the fact the manifold is compact 
in this direction) implies a discretization of the corresponding spectrum.

This difference in the physical content explains why do we find a 
field spectrum that is discrete in contrast to the case of 
absorption by branes where it is continuous. 
The important feature in this compactification that we are considering is that it 
makes it possible to map degrees of freedom of bulk and boundary theories
holographically\cite{BB}.

\section{Quantized fields in the Randall Sundrum scenario}
\label{RS}

Now we will discuss the implications of the results of section {\bf 4} 
about quantum fields in AdS space for the Randall Sundrum model.
First, we can change from the variable $z$ to $y$ 
defined by $ z\,=\, \Lambda \exp\{ y / \Lambda\} $. 
The metric (\ref{metric}) then takes the form 

\begin{equation}
\label{metric3}
ds^2=e^{-2ky}\eta_{\mu\nu}dx^\mu dx^\nu +dy^2\,,
\end{equation}

\noindent where $k = 1/\Lambda \,$. 

The Randall-Sundrum model\cite{RS1,RS2} corresponds to two 3-branes 
located at $y = 0$ and $y = r_c$, respectively, and the slice
of $AdS_5$ space between them (or also copies of it) 
with the background metric (\ref{metric3}), 
plus metric fluctuations.  
In this scenario the standard model fields live in the main
brane ($y = 0$) while gravity propagates in all dimensions including 
the fifth. 
Goldberger and Wise \cite{GW1}  studied the quantization of scalar 
fields in this model (in the AdS slice between the two branes).

In order to show the relation between the discussion on quantum fields in 
AdS space of section {\bf 4} and the proposal of Goldberger and Wise 
we can choose  $\delta = \Lambda$, {\sl i. e.}, now the main brane
is sitting on the beginning of one Poincare chart and the other brane 
is located at the end of the same chart ($R=\Lambda\exp\{r_c/\Lambda\}$).
The presence of the 3-branes in the Randall-Sundrum model leads to  
boundary conditions to be imposed on the bulk fields corresponding in 
our eqs. (\ref{BC1}),(\ref{BC2}) to $A = C = 0$.
In this case one obtains  solutions of the form of eq. (\ref{Sol1}), 
given in section {\bf 4}, but now the condition (\ref{Sol2}) 
that determines the eigenvalues $u_p$ is replaced by (in terms of the $z$ 
variable)

\begin{eqnarray}
&&\Big(2 J_{\nu}(u_p R ) + u_p R J^\prime_{\nu}(u_p R )\Big)
\Big( 2 Y_{\nu}(u_p \Lambda ) + 
u_p \Lambda Y^\prime_{\nu}(u_p \Lambda )\Big)\nonumber\\
&& - \Big( 2 Y_{\nu}(u_p R ) + u_p R Y^\prime_{\nu}(u_p R )\Big)
\Big( 2 J_{\nu}(u_p \Lambda ) + 
u_p \Lambda J^\prime_{\nu}(u_p \Lambda )\Big)\,\,=\,\,0\,.
\end{eqnarray}

\noindent  One could then repeat the discussion of section {\bf 4} 
following eq.(\ref{Sol2}) with the above condition. This would give 
the solutions found in \cite{GW1} written in terms of the coordinate 
$z$ and our parameters $R$ and $\Lambda$.

Once established this equivalence we can see what does our results  
on the field spectrum teaches us about the behavior of the model
when the second brane goes to infinity. 
The main brane accommodates the standard 
model fields and the observable physics lives there.
So one can think that we observe a projection of the extra dimension.
The existence of the second 
brane defines a compact AdS slice and this implies a discrete 
spectrum of bulk fields. As explained in \cite{GW1} this discretization 
makes a bulk scalar field looks like a tower of scalars for an observer on
the brane in a Kaluza Klein compactification mechanism. However,
if we take the coordinate $z$ (or $y$) as of
infinite range, removing the second brane, the spectrum would still 
be discrete.
This happens because, as studied in the previous
chapters, in this case we must use more than one coordinate chart.

%%%%%%%%%%%%%%%%%%%%%%%%%%%%%%%%%%%%%%%%%%%%%%%%%%%%%%%%%%%%%%%%%%%%%
%%%%%%%%%%%%%%%%%%%%%%%%%%%%%%%%%%%%%%%%%%%%%%%%%%%%%%%%%%%%%%%%%%%%%   
 
\section{Concluding Remarks}

We have seen that when we use Poincare coordinates to describe 
quantum fields in AdS space we must be careful about the fact that
one can not map the whole compactified space into just one chart.
This explains why the field spectrum in Poincare coordinates is also 
discrete despite the infinite coordinate ranges.
This discretization of the spectrum makes it possible 
to define a one to one correspondence between the degrees of 
freedom of the bulk and the boundary. 
We can understand this if we realize for example that the phase space
of the $AdS_5$  will correspond to a series of tridimensional hypersurfaces
that can be mapped into the tridimensional phase space of fields on 
the boundary.
If the spectrum were not discrete, the bulk phase space would correspond
 to a four dimensional manifold that would not map into the boundary 
phase space, violating the holographic principle.

As a remark, we note that we have only discussed the problem of 
the compactification of the axial coordinate $z$ because, as we saw here, 
it is essentially 
related to the possibility of mapping bulk and boundary degrees of freedom.
However, if we want a complete one to one mapping among Poincare coordinates 
and global coordinates of the compactified AdS space we should also consider
the compactification of the $x^i$ coordinates as well.
This happens because given some fixed finite $z$ and taking the limit 
$x^i \rightarrow \infty $  (or $ t \rightarrow \infty $) 
we also reach the boundary. 
This would require introducing more charts and imply that the 
field spectrum would be completely discrete.

Regarding the Randall Sundrum model we have seen that even if 
we remove the second brane we should introduce a second coordinate chart
which still implies a discrete field spectrum 
in the axial direction.
Then, for an observer in the main brane the bulk field would still 
effectively be represented as  a tower of fields. 
This mechanism can be though as a realization of the holographic
principle in the Randall Sundrum model.

In conclusion regarding $z$ as a coordinate associated with the 
renormalization group scale\cite{Ver,Ran} we can think that
the choice of the end of the Poincare chart could correspond to
the introduction of an energy scale.

%\vfill

%%%%%%%%%%%%%%%%%%%%%%%%%%%%%%%%%%%%%%%%%%%%%%%%%%%%%%%%%%%%%%%%%%%%%
%%%%%%%%%%%%%%%%%%%%%%%%%%%%%%%%%%%%%%%%%%%%%%%%%%%%%%%%%%%%%%%%%%%%%   

\section*{Acknowledgments} 
The authors were partially supported by CNPq, FINEP , FUJB and FAPERJ
- Brazilian research agencies. We also thank  Regina Celia Arcuri
and Franciscus Vanhecke for important discussions.

%%%%%%%%%%%%%%%%%%%%%%%%%%%%%%%%%%%%%%%%%%%%%%%%%%%%%%%%%%%%%%%%%%%%%%%%%

%\newpage

\end{document}